\documentclass[letterpaper, 10 pt, conference]{ieeeconf}  

\usepackage{cite}
\usepackage{amsmath,amssymb,amsfonts}
\usepackage{algorithmic}
\usepackage{graphicx}
\usepackage{textcomp}
\usepackage{xcolor}
\usepackage{siunitx}

\newtheorem{lemma}{Lemma}

\newtheorem{definition}{Definition}
\newtheorem{proposition}{Proposition}

\allowdisplaybreaks

\IEEEoverridecommandlockouts                              

\title{\LARGE \bf
Probabilistic Trajectory GOSPA: A Metric for Uncertainty-Aware\\ Multi-Object Tracking Performance Evaluation
}

\author{Yuxuan Xia$^{1}$, {\'A}ngel F. Garc{\'i}a-Fern{\'a}ndez$^{2}$, Johan Karlsson$^{3}$, Yu Ge$^{4}$, Lennart Svensson$^{4}$, Ting Yuan$^{1}$
\thanks{$^{1}$Yuxuan Xia and Ting Yuan are with the Department of Automation and Intelligent Sensing, Shanghai Jiao Tong University, Shanghai 200240, China. {\{yuxuan.xia,tyuan\}@sjtu.edu.cn}}%
\thanks{$^{2}${\'A}ngel F. Garc{\'i}a-Fern{\'a}ndez is with the Information Processing and Telecommunications Center, Universidad Politécnica de Madrid, Madrid, Spain. {angel.garcia.fernandez@upm.es}}%
\thanks{$^{3}$Johan Karlsson is with the Department of Mathematics, KTH Royal Institute of Technology, Stockholm, Sweden. {johan.karlsson@math.kth.se}}%
\thanks{$^{4}$Yu Ge and Lennart Svensson are with the Department of Electrical Engineering, Chalmers University of Technology, Gothenburg, Sweden. {\{yuge,lennart.svensson\}@chalmers.se}}%
\thanks{This work was partially funded by SJTU-KTH collaboration and research
development seed grants. (Corresponding author: Ting Yuan.)}
}

\begin{document}

\maketitle
\thispagestyle{empty}
\pagestyle{empty}

\begin{abstract}

This paper presents a generalization of the trajectory general optimal sub-pattern assignment (GOSPA) metric for evaluating multi-object tracking algorithms that provide trajectory estimates with track-level uncertainties. This metric builds on the recently introduced probabilistic GOSPA metric to account for both the existence and state estimation uncertainties of individual object states. Similar to trajectory GOSPA (TGOSPA), it can be formulated as a multidimensional assignment problem, and its linear programming relaxation---also a valid metric---is computable in polynomial time. Additionally, this metric retains the interpretability of TGOSPA, and we show that its decomposition yields intuitive costs terms associated to expected localization error and existence probability mismatch error for properly detected objects, expected missed and false detection error, and track switch error. The effectiveness of the proposed metric is demonstrated through a simulation study.

\end{abstract}

\section{Introduction}
The main objective of multiple object tracking (MOT) is to estimate the trajectories of moving objects, which may enter or leave the sensor's field-of-view, given noisy sensor measurements \cite{blackman1999design}. When developing and evaluating different MOT algorithms in different scenarios, it is very important to assess and compare their tracking performance using a suitable distance measure that quantifies the estimation error between the ground truth and the estimates.

A natural and minimal representation of the ground truth and the estimates is a set of trajectories, where a trajectory is a sequence of object states with a start time. To facilitate principled evaluation and intuitive error interpretation, it is desirable for the distance measure between two sets of trajectories to be a mathematically well-defined metric. However, many distance measures designed for sets of trajectories, e.g., \cite{bernardin2008evaluating,luiten2021hota,ristic2011metric}, are not metrics. Additionally, one would expect that the metric can penalize important aspects of MOT, including localization error for properly detected objects, missed and false detections, and track switches \cite{blackman1999design,fridling1991performance,drummond1992ambiguities}. Here, track switches refer to the change of optimal assignments of object states between true and estimated trajectories at different time steps \cite{blackman1999design}.

For the related problem of multiple object filtering, which estimates the set of object states at the current time step, two commonly used performance metrics are the optimal sub-pattern assignment (OSPA) metric \cite{schuhmacher2008new,schuhmacher2008consistent} and the generalized OSPA (GOSPA) metric \cite{rahmathullah2017generalized}. The OSPA metric penalizes cardinality mismatch, but it does not penalize missed and false detections based on intuitive concepts. As a comparison, the GOSPA metric penalizes localization errors for properly detected objects as well as missed and false detection errors, as typically required in MOT performance evaluation. Additionally,
GOSPA eliminates the spooky effect observed in optimal multi-object estimation using OSPA \cite{garcia2019spooky}, and it has
shown advantages over OSPA in sensor management \cite{garcia2021analysis}.

Both OSPA and GOSPA have been generalized to sets of trajectories for MOT performance evaluation. The OSPA$^{(2)}$ metric \cite{beard2017ospa} is such an extension of OSPA, which establishes an assignment between true and estimated trajectories that is not allowed to change over time. Thus, OSPA$^{(2)}$ cannot penalize track switch errors as desired in MOT performance evaluation \cite{blackman1999design}. A metric that generalizes GOSPA to sets of trajectories was proposed in \cite{garcia2020metric}, and it is usually referred to as the trajectory GOSPA (TGOSPA) metric \cite{krejvci2024tgospa}. In addition to penalizing localization errors and missed/false detections, TGOSPA also penalizes track switch errors based on changes in assignments and unassignments. Furthermore, it has been extended to incorporate weighting schemes that allow uneven penalization of costs across different time steps \cite{garcia2021time}.

For Bayesian MOT algorithms, object state estimates are extracted from the multi-object posterior at each time step. Then the MOT performance is often evaluated by computing the error between the ground truth and the state estimates. To account for the uncertainty information in the multi-object posterior densities, GOSPA has recently been extended to the space of multi-Bernoulli (MB) densities in \cite{xia2024probabilistic}, where each Bernoulli component is characterized by an existence probability and a single object density\footnote{The set of true object states can also be regarded as an MB density, where all the Bernoulli components have
existence probability one and Dirac delta single object densities.}. The resulting metric is termed probabilistic GOSPA (PGOSPA), since it enables a principled evaluation of multi-object filtering performance in probabilistic contexts.

\begin{figure}[!t]
\centerline{\includegraphics[width=0.8\linewidth]{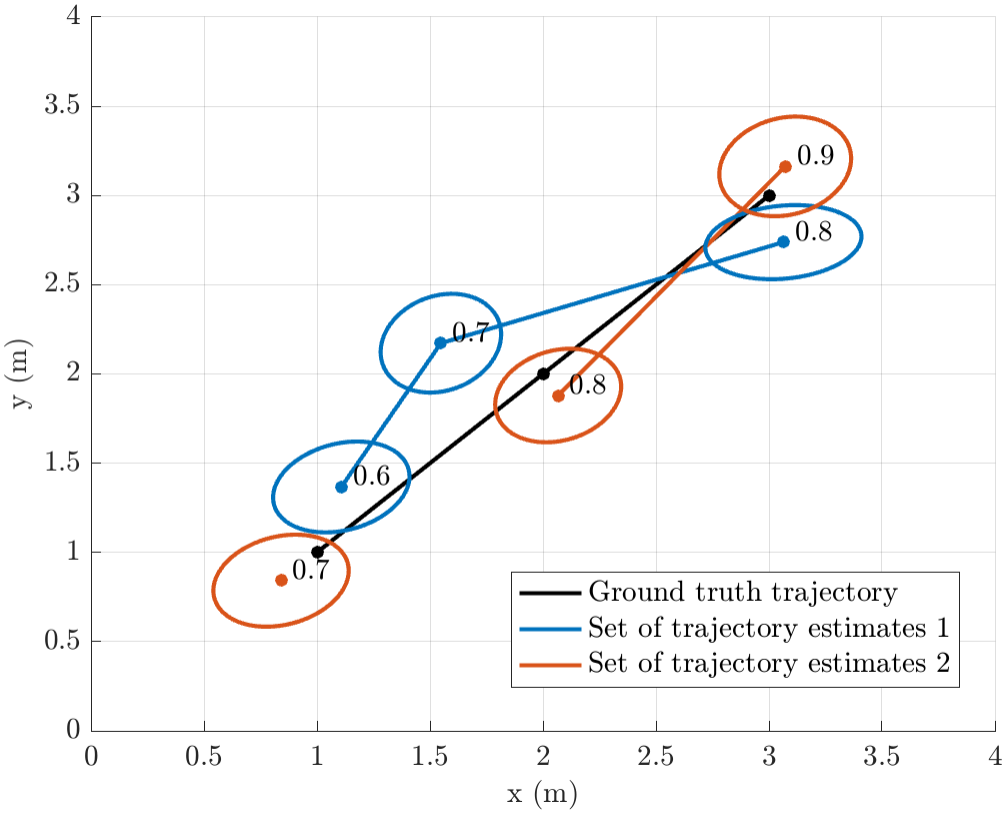}}
\caption{An exemplary scenario with a single ground truth trajectory and two sets of trajectory estimates, where each trajectory estimate is a sequence of Bernoulli densities. Each Bernoulli density has Gaussian single object density, and its existence probability is shown next to its Gaussian mean. The true trajectory and the trajectory estimate in blue exist at time step 1, 2 and 3. The set of trajectory estimates in orange consists of two single trajectory estimates, one exists only at time step 1, and the other exists at time step 2 and 3. A desirable metric should be able to answer: 1) what is the distance between each set of sequences of Bernoulli densities and the set of true trajectories? and 2) what is the distance between the two sets of sequences of Bernoulli densities?}
\label{fig_ex1}
\vspace{-5mm}
\end{figure}

Many MOT algorithms can report trajectory estimates with track-level uncertainties, e.g., the joint integrated probabilistic data association filter \cite{musicki2004joint}, the labelled MB filter \cite{reuter2014labeled}, and multi-object filters based on sets of trajectories \cite{garcia2019multiple,granstrom2025poisson,xia2019multi,garcia2020trajectory,garcia2020trajectory2}. To enable an uncertainty-aware performance evaluation for these MOT algorithms, it is desirable to have a metric that captures uncertainty information at the trajectory level, as illustrated in Fig. \ref{fig_ex1}. To this end, we extend TGOSPA to the space of sets of time sequences of Bernoulli densities by employing PGOSPA as the distance metric between Bernoulli densities. The resulting metric is referred to as probabilistic TGOSPA (PTGOSPA). 

The proposed PTGOSPA metric retains the interpretability of TGOSPA and PGOSPA. Furthermore, it admits a decomposition into five intuitive components: expected localization error and existence probability mismatch error for properly detected objects, expected missed and false detection error, and track switch error. Simulation results demonstrate that PTGOSPA provides richer insights into MOT performance than TGOSPA, by incorporating uncertainty information at the track level.


The rest of the paper is organized as follows. The background is introduced in Section II. The PTGOSPA metric is presented in Section III. Illustrative examples and simulation results are shown in Section IV. Conclusions are summarized in Section V.

\section{Background}

In this section, we first introduce the notation for sets of objects and sets of trajectories. Then we review the GOSPA and PGOSPA metrics, which are the building blocks for the TGOSPA and PTGOSPA metrics, respectively.

The single object state $x\in\mathbb{X}$ belongs to space $\mathbb{X}$, which is locally compact, Hausdorff and
second-countable \cite{mahler2007statistical}. A set $\mathbf{x}$ with $n_{\mathbf{x}}$ such elements is denoted $\mathbf{x} = \{x_1,\dots,x_{n_{\mathbf{x}}}\}\in\mathcal{F}(\mathbb{X})$, where $\mathcal{F}(\mathbb{X})$ denotes the set of finite subsets of $\mathbb{X}$. A single object trajectory is represented by $X=(t,x^{1:v})$, where $t$ is its start time, and $x^{1:v}$ is a sequence of $v$ object states at consecutive time steps. For trajectories that have existed in time interval $1:K$, $(t,v)$ belongs to the set $I_{(K)} = \{(t,v): 1\leq t\leq K, 1\leq v \leq K-t+1\}$. Then trajectory $X \in \uplus_{(t,v)\in I_{(K)}}\times \mathbb{X}^{v}$, where $\uplus$ denotes the union of mutually disjoint sets. A set of trajectories up to time step $K$ is denoted by $\mathbf{X} = \{X_1,\dots,X_{n_{\mathbf{X}}}\} \in \mathcal{F}(\uplus_{(t,v)\in I_{(K)}}\times \mathbb{X}^{v})$. 


\subsection{GOSPA metric}

We review the GOSPA metric with $\alpha=2$, which is the most appropriate choice for performance evaluation of MOT \cite{rahmathullah2017generalized}. We consider the sets of objects $\mathbf{x} = \{x_1,\dots,x_{n_\mathbf{x}}\}$ and $\mathbf{y} = \{y_1,\dots,y_{n_\mathbf{y}}\}$, and we let $\theta \in \Gamma_{\mathbf{x},\mathbf{y}}$ be an assignment set between $\{1,\dots,n_{\mathbf{x}}\}$ and $\{1,\dots,n_{\mathbf{y}}\}$ with the following properties: $\theta \subseteq \{1,\dots,n_{\mathbf{x}}\} \times \{1,\dots,n_{\mathbf{y}}\}$, $(i,j),(i,j^\prime)\in \theta \Rightarrow j = j^\prime$, and $(i,j),(i^\prime,j)\in \theta \Rightarrow i = i^\prime$, where $\Gamma_{\mathbf{x},\mathbf{y}}$ represents the set of all possible assignment sets. 

\begin{definition}
        Given a base metric $\bar{d}_b(x,y)$ between single object states $x,y\in\mathbb{X}$, two scalars $c>0$ and $1\leq p < \infty$, the GOSPA metric ($\alpha=2$) between the sets $\mathbf{x}$ and $\mathbf{y}$ can be formulated as an assignment problem over assignment sets \cite[Prop. 1]{rahmathullah2017generalized}:
        \begin{multline}
                {d}_G(\mathbf{x},\mathbf{y}) =\\ \min_{\theta\in\Gamma_{\mathbf{x},\mathbf{y}}}\left[ \sum_{(i,j)\in\theta} \bar{d}_b(x_i,y_j)^p + \frac{c^p}{2}(n_{\mathbf{x}} + n_{\mathbf{y}} - 2|\theta|) \right]^{\frac{1}{p}},
                \label{eq_gospa}
        \end{multline}
        where $|\theta|$ denotes the set cardinality of $\theta$, and the dependence of ${d}_G(\cdot,\cdot)$ on $p$ and $c$ has been dropped for notational brevity.
\end{definition} 

In GOSPA, $c$ is the cut-off distance representing the maximum allowable distance between two single object states, and larger values of $p$ impose stronger penalties on outliers. In \eqref{eq_gospa}, the first term corresponds to the localization error for assigned object states $x_i$ and $y_j$, with $(i,j)\in\theta$. The second term in \eqref{eq_gospa} can be further decomposed into missed detection error $\frac{c^p}{2}(n_{\mathbf{x}}-|\theta|)$ and false detection error $\frac{c^p}{2}(n_{\mathbf{y}}-|\theta|)$ (by assuming that $\mathbf{x}$ and $\mathbf{y}$ represents the true and estimated sets of object states, respectively).

In the case where both sets $\mathbf{x}$ and $\mathbf{y}$ contain at most one element, i.e., $n_\mathbf{x} \leq 1$ and $n_\mathbf{y} \leq 1$, \eqref{eq_gospa} reduces to
\begin{equation}
        {d}_G(\mathbf{x},\mathbf{y}) = \begin{cases}
                \min\left(\bar{d}_b(x,y),c\right) & \mathbf{x} = \{x\}, \mathbf{y} = \{y\}\\
                0 & \mathbf{x} = \mathbf{y} = \emptyset \\
                \frac{c}{2^{1/p}} & \text{otherwise},
        \end{cases}
\end{equation}
which is used in the TGOSPA metric \cite{garcia2020metric}.

\subsection{PGOSPA metric}

We first introduce the MB density on which the PGOSPA metric is evaluated. A Bernoulli process $\mathbf{x}$ is a set with a Bernoulli-distributed cardinality $|\mathbf{x}|$, and its density is 
\begin{equation}
        f(\mathbf{x}) = \begin{cases}
                1-r & \mathbf{x} = \emptyset \\
                rp(x) & \mathbf{x} = \{x\} \\
                0 & \text{otherwise},
        \end{cases}
\end{equation}
where $r\in[0,1]$ is the existence probability, and $p(\cdot)$ is the single object density conditioned on object existence. An MB process with $n$ Bernoulli components is a disjoint union of $n$ independent Bernoulli sets. Let the $i$-th Bernoulli component be parameterized by $r_i$ and $p_i(\cdot)$ where $i\in\{1,\dots,n\}$. Then the density of the MB process can be completely described by parameters $\{(r_i,p_i(\cdot))\}_{i=1}^n$.

Let $f_\mathbf{x}(\cdot)$ and $f_\mathbf{y}(\cdot)$ be two MB densities, parameterized by $\{(r^x_i,p^x_i(\cdot))\}_{i=1}^{n_\mathbf{x}}$ and $\{(r^y_i,p^y_i(\cdot))\}_{i=1}^{n_\mathbf{y}}$, respectively, where $r^x_i>0,r^y_i>0~\forall~ i$. We assume that $f_\mathbf{x}(\cdot)$ and $f_\mathbf{y}(\cdot)$ represent the true and estimated multi-object densities, respectively.

\begin{definition}
        Given a base metric $d_b(p^x,p^y)$ between single object densities $p^x(\cdot)$ and $p^y(\cdot)$, $c>0$ and $1\leq p < \infty$, the PGOSPA metric ($\alpha=2$) between MB densities $f_\mathbf{x}(\cdot)$ and $f_\mathbf{y}(\cdot)$ can be formulated as an assignment problem over assignment sets \cite[Prop. 2]{xia2024probabilistic}:
        \begin{align}
        &d_P(f_\mathbf{x},f_\mathbf{y})=\nonumber\\ 
        & \left[\min_{\theta\in\Gamma_{\mathbf{x},\mathbf{y}}}\left(\sum_{(i,j)\in\theta}\left[\min\left(r^x_i,r^y_{j}\right)d_b\left(p^x_i,p^y_{j}\right)^p + \left|r^x_i-r^y_j\right|\frac{c^p}{2} \right] \right.\right.\nonumber \\
        &~~~~~\left.\left.+\frac{c^p}{2}\left(\sum_{i:\forall j,(i,j)\notin\theta}r^x_i + \sum_{j:\forall i,(i,j)\notin\theta}r^y_j\right) \right)\right]^{\frac{1}{p}},
        \label{eq_pgospa}
    \end{align}
    where the dependence of $d_P(\cdot,\cdot)$ on $p$ and $c$ is dropped for notational brevity.
\end{definition}

The different terms in \eqref{eq_pgospa} can be interpreted as follows:
\begin{itemize}
    \item $\sum_{(i,j)\in\theta}\min(r^x_i,r^y_j)d(p^x_i,p^y_j)^p$: the expected localization error for associated Bernoulli components;
    \item $\sum_{(i,j)\in\theta}|r^x_i-r^y_j|c^p/2$: the existence probability mismatch error for associated Bernoulli components;
    \item $c^p/2\sum_{i:\forall j,(i,j)\notin\theta}r^x_i$: the expected missed detection error;
    \item $c^p/2\sum_{j:\forall i,(i,j)\notin\theta}r^y_j$: the expected false detection error.
\end{itemize}
In the expected localization error, the term $\min(r^x_i,r^y_j)$ arises from the $p$-Wasserstein distance between two Bernoulli set densities, as shown in \cite[Lemma 2]{xia2024probabilistic}. When both $f_\mathbf{x}(\cdot)$ and $f_\mathbf{y}(\cdot)$ have a single Bernoulli component, parameterized by $(r^x,p^x(\cdot))$ and $(r^y,p^y(\cdot))$, respectively, \eqref{eq_pgospa} reduces to
\begin{multline}
        d_P(f_\mathbf{x},f_\mathbf{y}) =\\ \left(\min(r^x,r^y)\min(d_b(p^x,p^y),c)^p + |r^x-r^y|\frac{c^p}{2}\right)^{\frac{1}{p}},
        \label{eq_pgospa_ber}
\end{multline}
which will be used in the PTGOSPA metric.

\section{Probabilistic Trajectory GOSPA metric}

In this section, we first introduce the PTGOSPA metric formulated using multidimensional assignment. We then present its linear programming relaxation based on soft assignments. Finally, we demonstrate how PTGOSPA can be decomposed into intuitive cost components.

\subsection{Notation}

Let $\mathcal{P}(\mathbb{X})$ be the space of probability densities on single object state space $\mathbb{X}$. Then the space of Bernoulli densities is $\mathbb{B} = [0,1]\times \mathcal{P}(\mathbb{X})$. A time sequence of Bernoulli densities is represented by $\mathsf{X} = (t,\mathsf{x}^{1:v})$, where $t$ is the start time step, $\mathsf{x}^{1:v}$ is a sequence of $v$ Bernoulli densities at consecutive time steps. We refer to the existence probability and single object density of the $i$-th Bernoulli density $\mathsf{x}^i$, $i\in\{1,\dots,v\}$, using $\mathsf{x}^i(r)$ and $\mathsf{x}^i(p)$, respectively.

The space of time sequences of Bernoulli densities in time interval $1:K$ is given by $\uplus_{(t,v)\in I_{(K)}}\times \mathbb{B}^{v}$. A set of time sequences of Bernoulli densities is $\boldsymbol{\mathsf{X}} = \{\mathsf{X}_1,\dots,\mathsf{X}_{n_{\boldsymbol{\mathsf{X}}}}\}$, and its space is $\mathcal{F}(\uplus_{(t,v)\in I_{(K)}}\times \mathbb{B}^{v})$. The PTGOSPA metric $d: \mathcal{F}(\uplus_{(t,v)\in I_{(K)}}\times \mathbb{B}^{v}) \times \mathcal{F}(\uplus_{(t,v)\in I_{(K)}}\times \mathbb{B}^{v}) \rightarrow \mathbb{R}_{\geq 0}$ measures the discrepancy between sets of Bernoulli sequences, whose start and end time steps are within time interval $1:K$.

Let $\boldsymbol{\mathsf{X}} = \{\mathsf{X}_1,\dots,\mathsf{X}_{n_{\boldsymbol{\mathsf{X}}}}\}$ and $\boldsymbol{\mathsf{Y}} = \{\mathsf{Y}_1,\dots,\mathsf{Y}_{n_{\boldsymbol{\mathsf{Y}}}}\}$ be two sets of Bernoulli sequences. Given Bernoulli sequence $\mathsf{X} = (t,\mathsf{x}^{1:v})$, the set of Bernoulli densities at time step $k$ is
\begin{equation}
        \tau^k(\mathsf{X}) = \begin{cases}
                \left\{\mathsf{x}^{k+v-1}\right\} & t \leq k \leq t + v - 1 \\
                \emptyset & \text{otherwise},
        \end{cases}
\end{equation}
and given a set $\boldsymbol{\mathsf{X}}$ of Bernoulli sequences, the set of Bernoulli densities at time step $k$ is
\begin{equation}
        \tau^k(\boldsymbol{\mathsf{X}}) = \bigcup_{\mathsf{X} \in \boldsymbol{\mathsf{X}}} \tau^k(\mathsf{X}).
\end{equation}
The sets of Bernoulli densities corresponding to Bernoulli sequences $\mathsf{X}_i$ and $\mathsf{Y}_j$ at time step $k$ are denoted by $\boldsymbol{\mathsf{x}}^k_i = \tau^k(\mathsf{X}_i)$ and $\boldsymbol{\mathsf{y}}^k_j = \tau^k(\mathsf{Y}_j)$, respectively.

In the PTGOSPA metric, at each time step $k$, each set $\boldsymbol{\mathsf{x}}^k_i$ can either be assigned to a set $\boldsymbol{\mathsf{y}}^k_j$ or be left unassigned. To account for the case of unassignment of $\boldsymbol{\mathsf{x}}^k_i$, we consider assignment between the sets $\{1,\dots,n_{\boldsymbol{\mathsf{X}}}\}$ and $\{0,\dots,n_{\boldsymbol{\mathsf{Y}}}\}$. Let $\pi^k = [\pi_1^k,\dots,\pi_{n_{\boldsymbol{\mathsf{X}}}}^k] \in \Pi_{\boldsymbol{\mathsf{X}},\boldsymbol{\mathsf{Y}}}$ be the assignment vector, where the set of all possible assignments is denoted by $\Pi_{\boldsymbol{\mathsf{X}},\boldsymbol{\mathsf{Y}}}$, and $\pi^k_i$ satisfies that $\pi^k_i = \pi^k_{i^\prime} = j > 0 \Rightarrow i = i^\prime$. This means that any two different $\boldsymbol{\mathsf{x}}^k_i$ cannot be assigned to the same $\boldsymbol{\mathsf{y}}^k_j$, but multiple Bernoulli sequences in $\boldsymbol{\mathsf{X}}$ and $\boldsymbol{\mathsf{Y}}$ can be left unassigned. We also let $\boldsymbol{\mathsf{X}}(\tilde{\pi}^k)$ and $\boldsymbol{\mathsf{Y}}(\tilde{\pi}^k)$ denote the subset of $\boldsymbol{\mathsf{X}}$ and $\boldsymbol{\mathsf{Y}}$ that are left unassigned according to $\pi^k$, respectively.


\subsection{PTGOSPA based on multidimensional assignment}

\begin{definition}
        Given a base metric $d_b(\cdot,\cdot)$ in the space of single object densities $\mathcal{P}(\mathbb{X})$, cut-off parameter $c>0$, $1\leq p < \infty$, and switching cost $\gamma > 0$, the PTGOSPA metric based on multidimensional assignment between sets $\boldsymbol{\mathsf{X}}$ and $\boldsymbol{\mathsf{Y}}$ of Bernoulli sequences is
        \begin{multline}
                d(\boldsymbol{\mathsf{X}},\boldsymbol{\mathsf{Y}}) = \\
                \min_{\substack{\pi^k\in\Pi_{\boldsymbol{\mathsf{X}},\boldsymbol{\mathsf{Y}}}\\k = 1,\dots,K}} \left[\sum_{k=1}^Kd^k_{\boldsymbol{\mathsf{X}},\boldsymbol{\mathsf{Y}}}\left(\boldsymbol{\mathsf{X}},\boldsymbol{\mathsf{Y}},\pi^k\right)^p + \sum_{k=1}^{T-1}s_{\boldsymbol{\mathsf{X}},\boldsymbol{\mathsf{Y}}}\left(\pi^k,\pi^{k+1}\right)^p\right]^{\frac{1}{p}}
                \label{eq_ptgospa}
        \end{multline}
        where the costs corresponding to properly detected objects, missed and false detections at time step $k$ are
        \begin{multline}
                d^k_{\boldsymbol{\mathsf{X}},\boldsymbol{\mathsf{Y}}}\left(\boldsymbol{\mathsf{X}},\boldsymbol{\mathsf{Y}},\pi^k\right)^p = \sum_{(i,j)\in\theta^k(\pi^k)}d_P\left(\boldsymbol{\mathsf{x}}^k_i,\boldsymbol{\mathsf{y}}^k_j\right)^p\\
                + \frac{c^p}{2}\left(\sum_{\mathsf{x}\in\tau^k(\boldsymbol{\mathsf{X}}(\tilde{\theta}^k(\pi^k)))}\mathsf{x}(r) + \sum_{\mathsf{y}\in\tau^k(\boldsymbol{\mathsf{Y}}(\tilde{\theta}^k(\pi^k)))}\mathsf{y}(r)\right),
                \label{eq_pgospa_p_power}
        \end{multline}
        with 
        \begin{multline}
                \theta^k(\pi^k) = \left\{(i,j): i \in \{1,\dots,n_{\boldsymbol{\mathsf{X}}}\},\right.\\\left. |\boldsymbol{\mathsf{x}}^k_i| = |\boldsymbol{\mathsf{y}}^k_j| = 1, d_b(\boldsymbol{\mathsf{x}}^k_i(p),\boldsymbol{\mathsf{y}}^k_j(p)) < c\right\},
                \label{eq_excluding_c}
        \end{multline}
        and the switching cost from time step $k$ to $k+1$ is given by
        \begin{align}
                s_{\boldsymbol{\mathsf{X}},\boldsymbol{\mathsf{Y}}}\left(\pi^k,\pi^{k+1}\right)^p &= \gamma^p\sum_{i=1}^{n_{\boldsymbol{\mathsf{X}}}}s(\pi^k_i,\pi^{k+1}_i),\label{eq_switch_cost}\\
                s(\pi^k_i,\pi^{k+1}_i) &= \begin{cases}
                        0 & \pi^k_i = \pi^{k+1}_i \\
                        1 & \pi^k_i \neq \pi^{k+1}_i, \pi^k_i \neq 0, \pi^{k+1}_i \neq 0 \\
                        1/2 & \text{otherwise}.
                \end{cases}
                \label{eq_switch_cost_decom}
        \end{align}
\end{definition}

Note that \eqref{eq_pgospa_p_power} can be regarded as the PGOSPA metric raised to the $p$-th power, but without the minimization, see \eqref{eq_pgospa} and \eqref{eq_pgospa_ber}. Instead of performing minimization over an object-level assignment $\theta^k$, the assignment is induced by the trajectory-level assignment $\pi^k$. In this formulation, for $(i,j)\in\theta^k(\pi^k)$, $\boldsymbol{\mathsf{x}}^k_i$ and $\boldsymbol{\mathsf{y}}^k_j$ each contain precisely one Bernoulli density, and the distance between their single object densities is less than $c$, see \eqref{eq_excluding_c}. Under this condition, it also follows from \eqref{eq_pgospa_ber} that the distance between the two Bernoulli densities satisfies 
\begin{equation}
        d_P\left(\boldsymbol{\mathsf{x}}^k_i,\boldsymbol{\mathsf{y}}^k_j\right) < c\left(\frac{\boldsymbol{\mathsf{x}}^k_i(r)+\boldsymbol{\mathsf{y}}^k_j(r)}{2}\right)^{\frac{1}{p}}.
\end{equation}
We can conclude that \eqref{eq_pgospa_p_power} represents the sum of the expected localization error and the existence probability mismatch error for properly detected objects (the first term), the expected missed detection error (the second term), and the expected false detection error (the third term).

The change from the assignment vector at the trajectory-level from $\pi^k$ to $\pi^{k+1}$ determines the track switching cost \eqref{eq_switch_cost}. The term $s(\pi^k_i,\pi^{k+1}_i)$ \eqref{eq_switch_cost_decom} corresponds to no switch when there are no changes in the assignments, a full switch when there is a change between two non-zero assignments, and a half switch when there is a change from a non-zero to zero assignment or vice versa \cite{garcia2020metric}. 

The TGOSPA metric is as a special case of the PTGOSPA metric where all the Bernoulli densities have existence probability one and Dirac delta single object densities. Therefore, PTGOSPA can be considered as a probabilistic generalization of TGOSPA. The proof that PTGOSPA is a mathematically well-defined metric follows analogously to that of TGOSPA.

\subsection{Linear programming relaxation of PTGOSPA}

The assignment vectors $\pi^k \in \Pi_{\boldsymbol{\mathsf{X}},\boldsymbol{\mathsf{Y}}}$ in \eqref{eq_ptgospa} can be equivalently represented using binary weight matrices. Let $\mathcal{W}_{\boldsymbol{\mathsf{X}},\boldsymbol{\mathsf{Y}}}$ be the set of all binary matrices of dimension $(n_{\boldsymbol{\mathsf{X}}}+1)\times(n_{\boldsymbol{\mathsf{Y}}}+1)$, representing assignments between $\boldsymbol{\mathsf{X}}$ and $\boldsymbol{\mathsf{Y}}$. A matrix $W^k \in \mathcal{W}_{\boldsymbol{\mathsf{X}},\boldsymbol{\mathsf{Y}}}$ satisfies the following properties \cite{garcia2020metric}:
\begin{subequations}
\begin{align}
        \sum_{i=1}^{n_{\boldsymbol{\mathsf{X}}}+1} W^k(i,j) &= 1, j = 1,\dots,n_{n_{\boldsymbol{\mathsf{Y}}}},\label{eq_binary_constraint1}\\
        \sum_{j=1}^{n_{\boldsymbol{\mathsf{Y}}}+1} W^k(i,j) &= 1, i = 1,\dots,n_{n_{\boldsymbol{\mathsf{X}}}},\label{eq_binary_constraint2}\\
        W^k(n_{\boldsymbol{\mathsf{X}}}+1,n_{\boldsymbol{\mathsf{Y}}}+1) &= 0,\label{eq_binary_constraint3}\\
        W^k(i,j) &\in \{0,1\}, \forall i,j,\label{eq_binary_constraint}
\end{align}
\end{subequations}
where $W^k(i,j)$ represents the element at the $i$-th row and $j$-th column of matrix $W^k$. The bijection between the sets $\Pi_{\boldsymbol{\mathsf{X}},\boldsymbol{\mathsf{Y}}}$ and $\mathcal{W}_{\boldsymbol{\mathsf{X}},\boldsymbol{\mathsf{Y}}}$ is given by 
\begin{align*}
        \pi^k_i = j \neq 0 &\Leftrightarrow W^k(i,j) = 1,\\
        \pi^k_i = 0 &\Leftrightarrow W^k(i,n_{\boldsymbol{\mathsf{Y}}}+1) = 1,\\
        \nexists i\in\{1,\dots,n_{\boldsymbol{\mathsf{X}}}\},\pi_i^k = j \neq 0 &\Leftrightarrow W^k(n_{\boldsymbol{\mathsf{X}}}+1,j) = 1,
\end{align*}
for $\pi^k \in \Pi_{\boldsymbol{\mathsf{X}},\boldsymbol{\mathsf{Y}}}$, $W^k \in \mathcal{W}_{\boldsymbol{\mathsf{X}},\boldsymbol{\mathsf{Y}}}$, $i = 1,\dots, n_{\boldsymbol{\mathsf{X}}}$, $j = 1,\dots, n_{\boldsymbol{\mathsf{Y}}}$.

\begin{lemma}
        The PTGOSPA metric $d(\boldsymbol{\mathsf{X}},\boldsymbol{\mathsf{Y}})$ in \eqref{eq_ptgospa} between sets $\boldsymbol{\mathsf{X}}$ and $\boldsymbol{\mathsf{Y}}$ of Bernoulli sequences can be formulated as the following binary linear program
        \begin{multline}
                d(\boldsymbol{\mathsf{X}},\boldsymbol{\mathsf{Y}}) = \min_{\substack{W^k \in \mathcal{W}_{\boldsymbol{\mathsf{X}},\boldsymbol{\mathsf{Y}}}\\k = 1,\dots, K}} \left[\sum_{k=1}^K\text{trace}\left(\left(D^k_{\boldsymbol{\mathsf{X}},\boldsymbol{\mathsf{Y}}}\right)^{\text{T}}W^k\right)\right. \\\left. +\frac{\gamma^p}{2}\sum_{k=1}^{K-1} \sum_{i=1}^{n_{\boldsymbol{\mathsf{X}}}}\sum_{j=1}^{n_{\boldsymbol{\mathsf{Y}}}} \left|W^k(i,j) - W^{k+1}(i,j)\right|\right]^{\frac{1}{p}},
                \label{eq_binary_lp}
        \end{multline}
        where $D^k_{\boldsymbol{\mathsf{X}},\boldsymbol{\mathsf{Y}}}$ is a $(n_{\boldsymbol{\mathsf{X}}}+1)\times(n_{\boldsymbol{\mathsf{Y}}}+1)$ matrix whose $(i,j)$-th element is 
        \begin{equation}
                D^k_{\boldsymbol{\mathsf{X}},\boldsymbol{\mathsf{Y}}}(i,j) = \begin{cases}
                        d_P\left(\boldsymbol{\mathsf{x}}^k_i,\boldsymbol{\mathsf{y}}^k_j\right)^p & \boldsymbol{\mathsf{x}}^k_i \neq \emptyset, \boldsymbol{\mathsf{y}}^k_i \neq \emptyset \\
                        \boldsymbol{\mathsf{x}}^k_i(r)\frac{c^p}{2} & \boldsymbol{\mathsf{x}}^k_i \neq \emptyset, \boldsymbol{\mathsf{y}}^k_i = \emptyset \\
                        \boldsymbol{\mathsf{y}}^k_j(r)\frac{c^p}{2} & \boldsymbol{\mathsf{x}}^k_i = \emptyset, \boldsymbol{\mathsf{y}}^k_i \neq \emptyset \\
                        0 & \boldsymbol{\mathsf{x}}^k_i = \emptyset, \boldsymbol{\mathsf{y}}^k_i = \emptyset,
                \end{cases}
        \end{equation}
        with $\boldsymbol{\mathsf{x}}^k_{n_{\boldsymbol{\mathsf{X}}}+1} = \emptyset$ and $\boldsymbol{\mathsf{y}}^k_{n_{\boldsymbol{\mathsf{Y}}}+1} = \emptyset$.
\end{lemma}

The binary constraint \eqref{eq_binary_constraint} can be relaxed to define the set $\overline{\mathcal{W}}_{\boldsymbol{\mathsf{X}},\boldsymbol{\mathsf{Y}}}$ of matrices, where each matrix $W^k \in \overline{\mathcal{W}}_{\boldsymbol{\mathsf{X}},\boldsymbol{\mathsf{Y}}}$ satisfies \eqref{eq_binary_constraint1}, \eqref{eq_binary_constraint2}, \eqref{eq_binary_constraint3} and 
\begin{equation}
        W^k(i,j) \geq 0, \forall i,j.\label{eq_binary_constraint_new}
\end{equation}
\begin{proposition}
        Given a base metric $d_b(\cdot,\cdot)$ in the space of single object densities $\mathcal{P}(\mathbb{X})$, cut-off parameter $c>0$, $1\leq p < \infty$, and switching cost $\gamma > 0$, the linear programming relaxation $\bar{d}(\boldsymbol{\mathsf{X}},\boldsymbol{\mathsf{Y}})$ of $d(\boldsymbol{\mathsf{X}},\boldsymbol{\mathsf{Y}})$ \eqref{eq_binary_lp} is also a metric, given by
        \begin{multline}
                \bar{d}(\boldsymbol{\mathsf{X}},\boldsymbol{\mathsf{Y}}) = \min_{\substack{W^k \in \overline{\mathcal{W}}_{\boldsymbol{\mathsf{X}},\boldsymbol{\mathsf{Y}}}\\k = 1,\dots, K}} \left[\sum_{k=1}^K\text{trace}\left(\left(D^k_{\boldsymbol{\mathsf{X}},\boldsymbol{\mathsf{Y}}}\right)^{\text{T}}W^k\right)\right. \\\left. +\frac{\gamma^p}{2}\sum_{k=1}^{K-1} \sum_{i=1}^{n_{\boldsymbol{\mathsf{X}}}}\sum_{j=1}^{n_{\boldsymbol{\mathsf{Y}}}} \left|W^k(i,j) - W^{k+1}(i,j)\right|\right]^{\frac{1}{p}},
                \label{eq_binary_lp_relax}
        \end{multline}
        where the matrix $D^k_{\boldsymbol{\mathsf{X}},\boldsymbol{\mathsf{Y}}}$ is defined as in \eqref{eq_binary_lp}.
\end{proposition}

The linear programming relaxation of PTGOSPA \eqref{eq_binary_lp_relax} is analogous to the linear programming relaxation of TGOSPA \cite[Eq. (23)]{garcia2020metric}, but with a different choice of $D^k_{\boldsymbol{\mathsf{X}},\boldsymbol{\mathsf{Y}}}$. Thus, the proof that the linear programming relaxation of PTGOSPA is also a mathematically well-defined metric follows analogously to that of TGOSPA \cite[App. B]{garcia2020metric}. 

With the relaxed constraint \eqref{eq_binary_constraint_new}, soft assignments between Bernoulli sequences are allowed, and the optimization is now performed over $W^k \in \overline{\mathcal{W}}_{\boldsymbol{\mathsf{X}},\boldsymbol{\mathsf{Y}}}$ instead of ${\mathcal{W}}_{\boldsymbol{\mathsf{X}},\boldsymbol{\mathsf{Y}}}$. As a result, the linear programming relaxation $\bar{d}(\boldsymbol{\mathsf{X}},\boldsymbol{\mathsf{Y}})$ of ${d}(\boldsymbol{\mathsf{X}},\boldsymbol{\mathsf{Y}})$ serves as a lower bound, i.e., $\bar{d}(\boldsymbol{\mathsf{X}},\boldsymbol{\mathsf{Y}}) \leq {d}(\boldsymbol{\mathsf{X}},\boldsymbol{\mathsf{Y}})$, and is computable in polynomial time.

\subsection{PTGOSPA decomposition}

We proceed to describe the decompositions of PTGOSPA formulated as a binary linear program and its relaxation via soft assignments \eqref{eq_binary_lp_relax}. We note that $d_P(\boldsymbol{\mathsf{x}}^k_i,\boldsymbol{\mathsf{y}}^k_j)^p$ represents the following costs (to the $p$-th power):
\begin{enumerate}
        \item The sum of an expected localization error 
        \begin{equation*}
                \min(\boldsymbol{\mathsf{x}}^k_i(r),\boldsymbol{\mathsf{y}}^k_j(r))d_b(\boldsymbol{\mathsf{x}}^k_i(p),\boldsymbol{\mathsf{y}}^k_j(p))^p,
        \end{equation*}
        and an existence probability mismatch error $|\boldsymbol{\mathsf{x}}^k_i(r)-\boldsymbol{\mathsf{y}}^k_j(r)|c^p/2$ for a properly detected object, if $i \leq n_{\boldsymbol{\mathsf{X}}}$, $j\leq n_{\boldsymbol{\mathsf{Y}}}$, $\boldsymbol{\mathsf{x}}^k_i \neq \emptyset$, $\boldsymbol{\mathsf{y}}^k_i \neq \emptyset$, $d_b(\boldsymbol{\mathsf{x}}^k_i(p),\boldsymbol{\mathsf{y}}^k_j(p)) < c$.
        \item An expected missed detection error $\boldsymbol{\mathsf{x}}^k_i(r)c^p/2$, if $i \leq n_{\boldsymbol{\mathsf{X}}}$, $j\leq n_{\boldsymbol{\mathsf{Y}}}$, $\boldsymbol{\mathsf{x}}^k_i \neq \emptyset$, $\boldsymbol{\mathsf{y}}^k_i = \emptyset$ or $i \leq n_{\boldsymbol{\mathsf{X}}}$, $j = n_{\boldsymbol{\mathsf{Y}}}+1$.
        \item An expected false detection error $\boldsymbol{\mathsf{y}}^k_j(r)c^p/2$, if $i \leq n_{\boldsymbol{\mathsf{X}}}$, $j\leq n_{\boldsymbol{\mathsf{Y}}}$, $\boldsymbol{\mathsf{x}}^k_i = \emptyset$, $\boldsymbol{\mathsf{y}}^k_i \neq \emptyset$ or $i = n_{\boldsymbol{\mathsf{X}}}+1$, $j \leq n_{\boldsymbol{\mathsf{Y}}}$.
        \item The sum of an expected missed detection error and an expected false detection error $(\boldsymbol{\mathsf{x}}^k_i(r)+\boldsymbol{\mathsf{y}}^k_j(r))c^p/2$, if $i \leq n_{\boldsymbol{\mathsf{X}}}$, $j\leq n_{\boldsymbol{\mathsf{Y}}}$, $\boldsymbol{\mathsf{x}}^k_i \neq  \emptyset$, $\boldsymbol{\mathsf{y}}^k_i \neq \emptyset$, $d_b(\boldsymbol{\mathsf{x}}^k_i(p),\boldsymbol{\mathsf{y}}^k_j(p)) = c$.
\end{enumerate}

We denote the sets of indices $(i,j)$ that belong to each of previous categories at time step $k$ as $T_1^k$, $T_2^k$, $T_3^k$ and $T_4^k$. Then, we can rewrite \eqref{eq_binary_lp} as 
\begin{align}
        d(\boldsymbol{\mathsf{X}},\boldsymbol{\mathsf{Y}}) &=\min_{\substack{W^k \in \mathcal{W}_{\boldsymbol{\mathsf{X}},\boldsymbol{\mathsf{Y}}}\\k = 1,\dots, K}}\left( \sum_{k=1}^K l^k(\boldsymbol{\mathsf{X}},\boldsymbol{\mathsf{Y}},W^k)\right. \nonumber\\
        &~~~+ \sum_{k=1}^K e^k(\boldsymbol{\mathsf{X}},\boldsymbol{\mathsf{Y}},W^k) + \sum_{k=1}^K m^k(\boldsymbol{\mathsf{X}},\boldsymbol{\mathsf{Y}},W^k) \nonumber\\
        &~~~+\left.\sum_{k=1}^K f^k(\boldsymbol{\mathsf{X}},\boldsymbol{\mathsf{Y}},W^k) + \sum_{k=1}^{K-1} s^k(\boldsymbol{\mathsf{X}},\boldsymbol{\mathsf{Y}},W^k)\right)^{\frac{1}{p}},
\end{align}
where the following costs (to the $p$-th power)
\begin{align*}
        l^k(\boldsymbol{\mathsf{X}},\boldsymbol{\mathsf{Y}},W^k)
        &= \sum_{(i,j)\in T^k_1}\min(\boldsymbol{\mathsf{x}}^k_i(r),\boldsymbol{\mathsf{y}}^k_j(r))\nonumber\\
        &~~~~~~~~~~~~\times d_b(\boldsymbol{\mathsf{x}}^k_i(p),\boldsymbol{\mathsf{y}}^k_j(p))^pW^k(i,j),\\                
        e^k(\boldsymbol{\mathsf{X}},\boldsymbol{\mathsf{Y}},W^k) &= \frac{c^p}{2}\sum_{(i,j)\in T^k_1}|\boldsymbol{\mathsf{x}}^k_i(r)-\boldsymbol{\mathsf{y}}^k_j(r)|W^k(i,j),\\
        m^k(\boldsymbol{\mathsf{X}},\boldsymbol{\mathsf{Y}},W^k) &= \frac{c^p}{2}\sum_{(i,j)\in T^k_2 \cup T^k_4} \boldsymbol{\mathsf{x}}^k_i(r)W^k(i,j),\\
        f^k(\boldsymbol{\mathsf{X}},\boldsymbol{\mathsf{Y}},W^k) &= \frac{c^p}{2}\sum_{(i,j)\in T^k_3 \cup T^k_4} \boldsymbol{\mathsf{y}}^k_j(r)W^k(i,j),\\ 
        s^k(\boldsymbol{\mathsf{X}},\boldsymbol{\mathsf{Y}},W^k) &= \frac{\gamma^p}{2}\sum_{i=1}^{n_{\boldsymbol{\mathsf{X}}}}\sum_{j=1}^{n_{\boldsymbol{\mathsf{Y}}}} \left|W^k(i,j) - W^{k+1}(i,j)\right|,
\end{align*}
represent 1) the expected localization error for properly detected objects, 2) the existence probability mismatch error for properly detected objects, 3) the expected missed detection error, 4) the expected false detection error, and 5) the track switch error, respectively. The linear programming relaxation of the PTGOSPA metric \eqref{eq_binary_lp_relax} has the same decomposition, but the assignments are soft.

\section{Simulation Results}

In this section, we compare the TGOSPA and PTGOSPA metrics by employing them to evaluate the MOT performance of the trajectory Poisson multi-Bernoulli mixture (TPMBM) filter \cite{granstrom2025poisson} and the trajectory Poisson multi-Bernoulli (TPMB) filter \cite{garcia2020trajectory}. The TPMBM filter provides a closed-form solution to the MOT problem under standard models for sets of trajectories with a Poisson point process (PPP) birth \cite{garcia2019multiple}. Its multi-object posterior density maintains the TPMBM form, representing the set of undetected trajectories with a PPP and the set of detected trajectories with a multi-Bernoulli mixture (MBM). By merging the MBM into a single MB after each update step, we obtain the TPMB filter \cite{garcia2020trajectory}.

For both TPMBM and TPMB, we consider the implementation that estimates the set of all trajectories using Gaussian information form \cite{granstrom2025poisson}. Both filters are implemented with the following parameters: ellipsoidal gating size 20, maximum number of global hypothesis 1000 (found using the Murty's algorithm \cite{crouse2016implementing}), threshold for pruning Bernoulli components $10^{-5}$, threshold for pruning Poisson intensity weights $10^{-5}$, threshold for not predicting/updating dead trajectories $10^{-5}$. In addition, for the TPMBM filter, the threshold for pruning global hypotheses is $10^{-4}$.

In both filters, each detected trajectory is represented by a trajectory Bernoulli density, parameterized by an existence probability and a single trajectory density. The time sequence of Bernoulli densities can be computed by marginalizing the trajectory Bernoulli density \cite[Theorem 4]{granstrom2025poisson}. Furthermore, for TPMBM, to fully capture the uncertainty in its MBM, we first compute the corresponding sets of Bernoulli sequences for each MB and then evaluate the PTGOSPA metric with respect to the ground truth, followed by taking their weighted sum (considering the global hypothesis weights). 

Performance evaluation using TGOSPA needs an estimator. For both filters, we report the most likely sequence of object state estimates by selecting Bernoulli components with existence probability greater than 0.5 from the MB with the highest weight only at the final time step. For both TGOSPA and PTGOSPA, we consider implementations based on linear programming relaxation and set $c=10$, $p=2$ and $\gamma=2$. In PTGOSPA, we use the 2-Wasserstein distance as the base metric between single object densities as in \cite{xia2024probabilistic}.

\begin{figure}[!t]
    \center
    \includegraphics[width=0.8\linewidth]{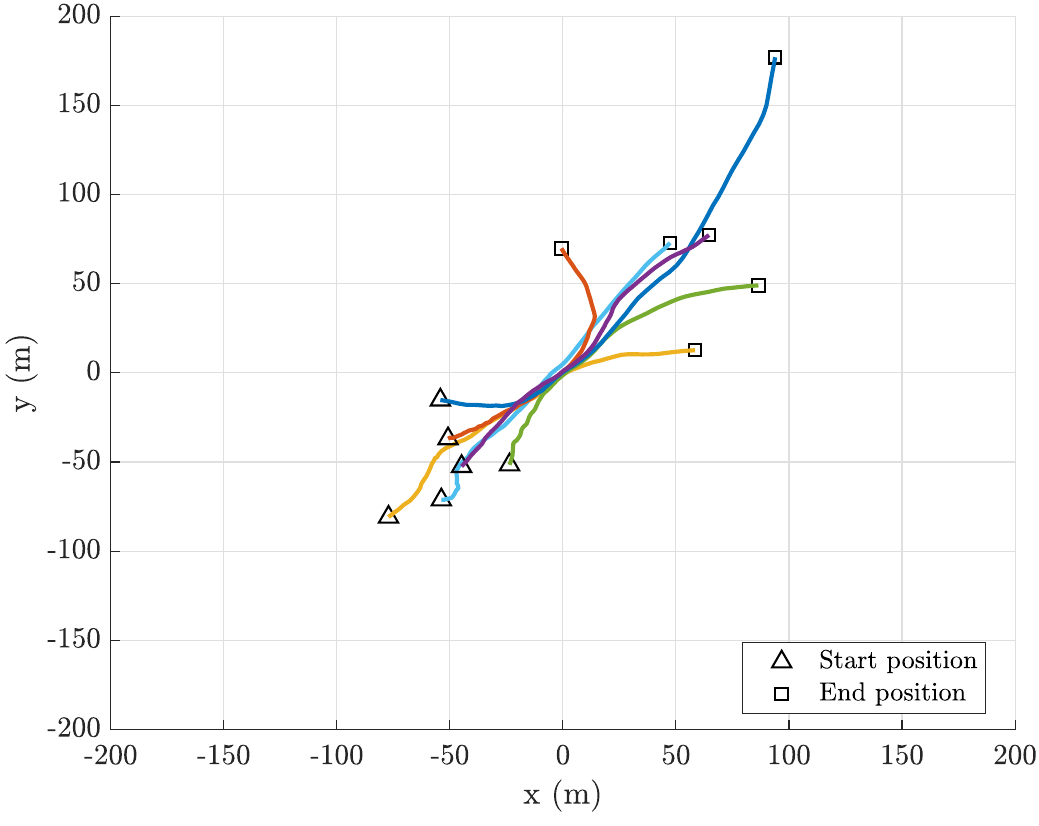}
    \caption{Ground truth of the simulated scenario, which give rise to a difficult data association problem. The start/end positions of trajectories are marked by triangles/squares. Six objects are born at time 1, 1, 11, 11, 21, and 21, and die at time 61, 61, 71, 71, 81, and 81, respectively.}
    \label{fig_gt}
    \vspace{-5mm}
\end{figure}

In the simulation, we adopt the same scenario as described in \cite{xia2024markov} where six initially separate objects move into close proximity before diverging again. The ground truth trajectories are illustrated in Fig. \ref{fig_gt}. Object states include 2D position and velocity, generated according to a PPP birth model with an intensity of 0.02. The birth density is modeled as a single Gaussian with a zero mean and a large covariance covering the entire area of interest. A nearly constant velocity motion model is used, with a sampling period of $\SI{1}{\second}$ and a process noise standard deviation of 0.3. Measurements follow a linear Gaussian model with an identity covariance matrix. We also consider an object survival probability of 0.98, a detection probability of 0.7, and Poisson clutter with a rate of 30 and uniform spatial distribution.

\begin{figure*}[!t]
    \center
    \includegraphics[width=0.32\linewidth]{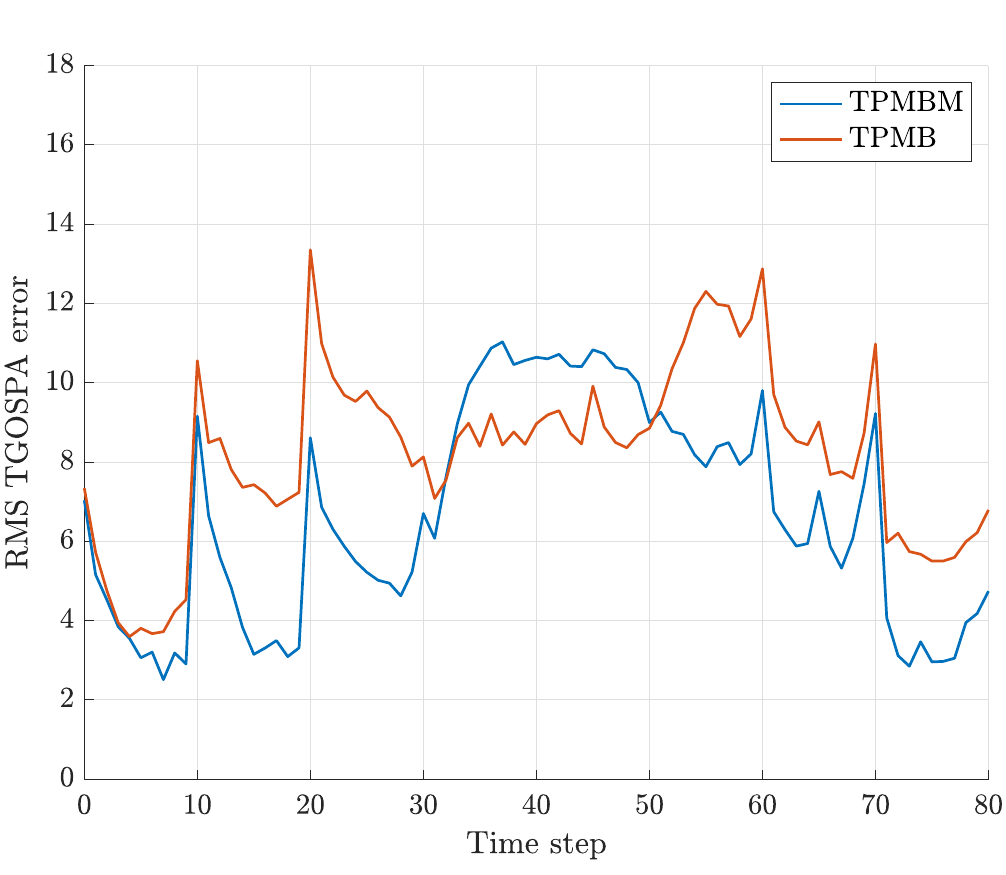}
    \includegraphics[width=0.32\linewidth]{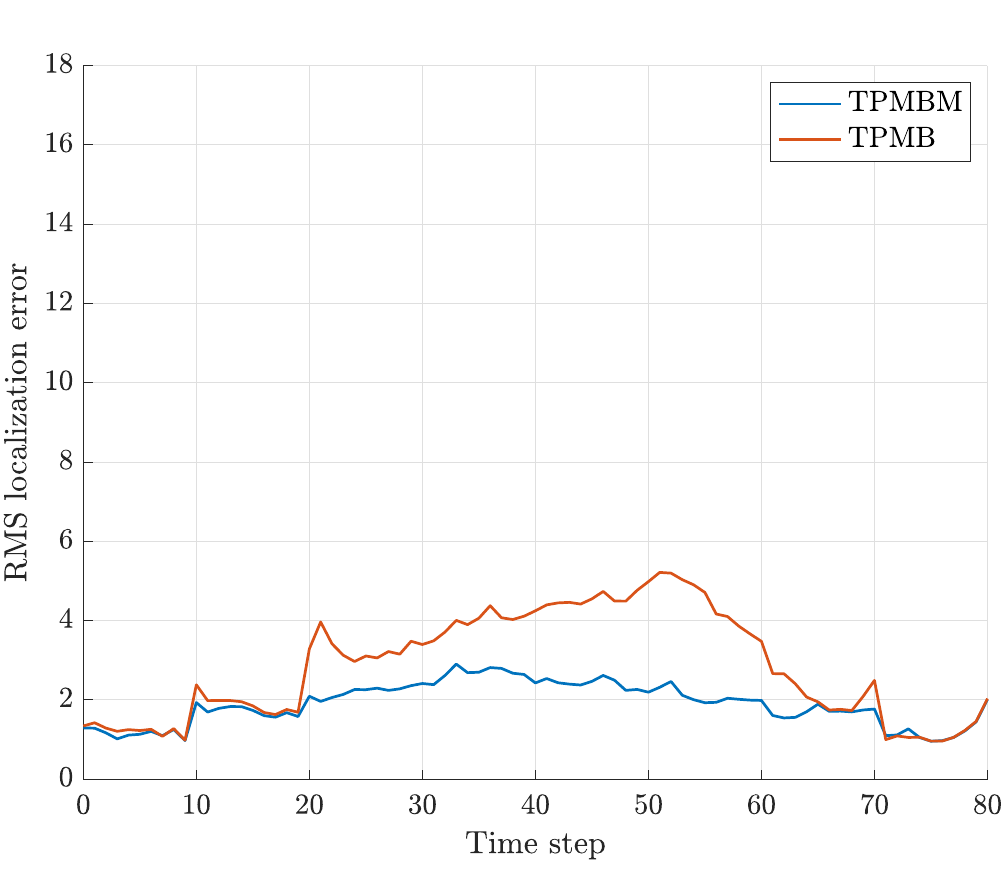}
    \includegraphics[width=0.32\linewidth]{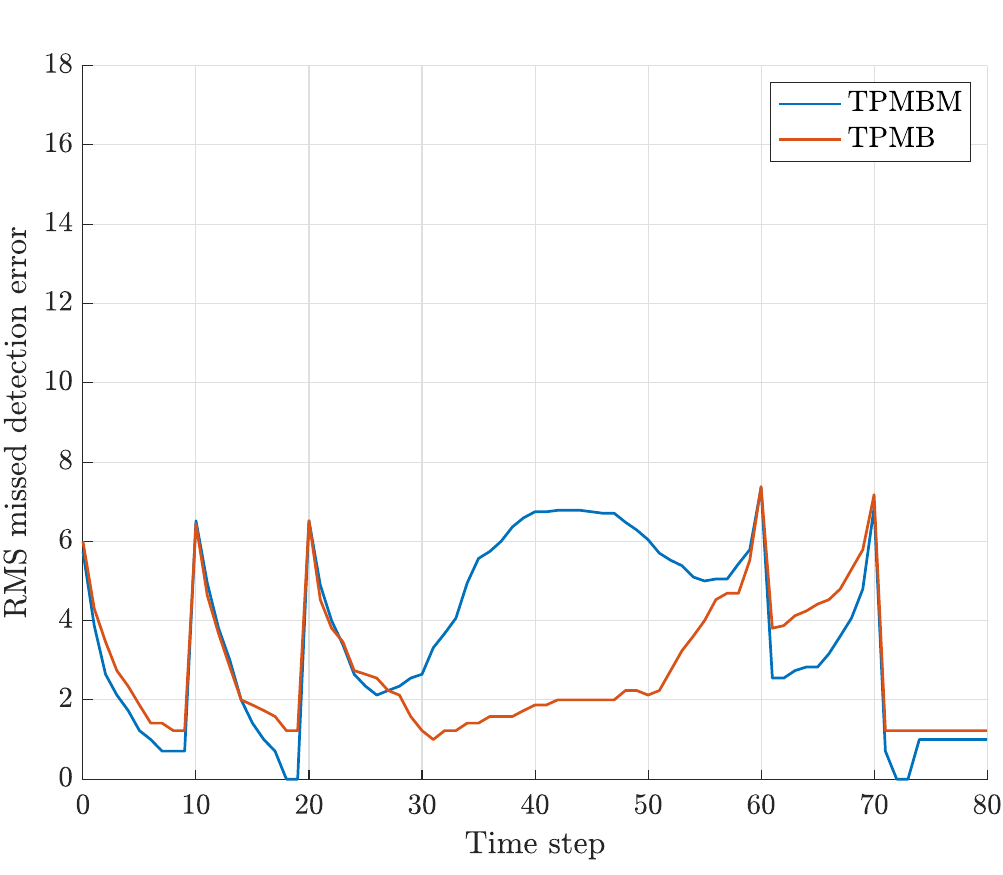}
    \includegraphics[width=0.32\linewidth]{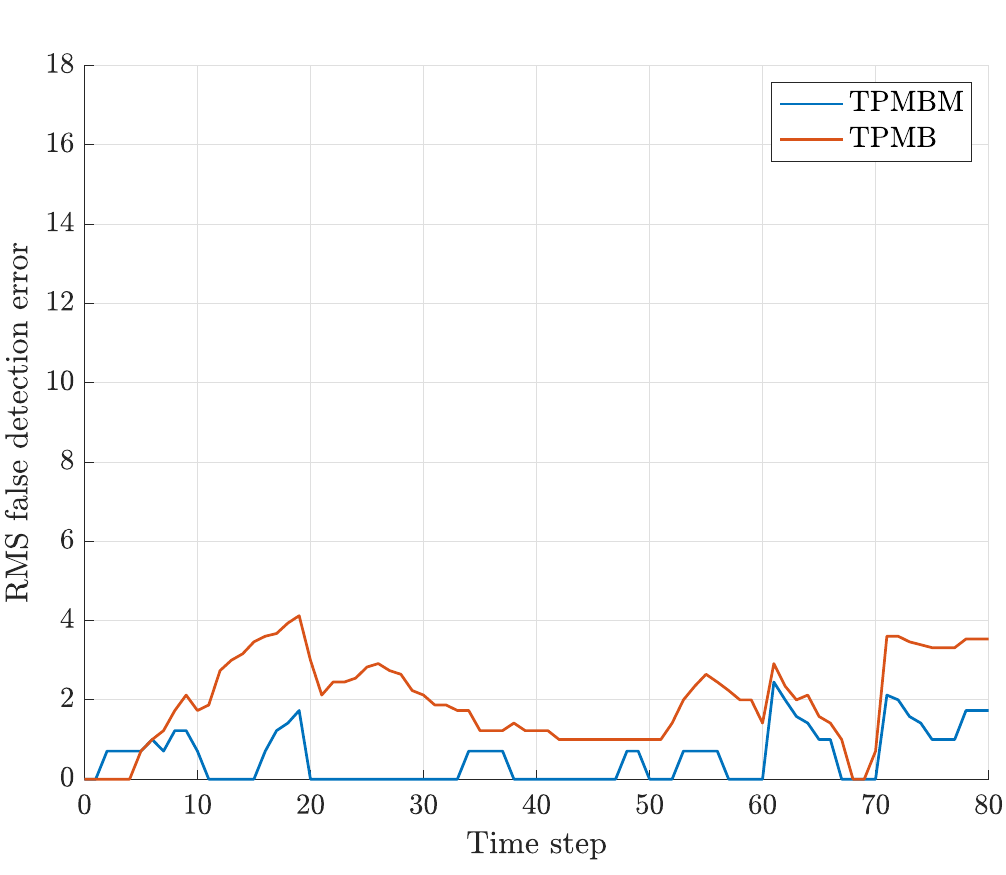}
    \includegraphics[width=0.32\linewidth]{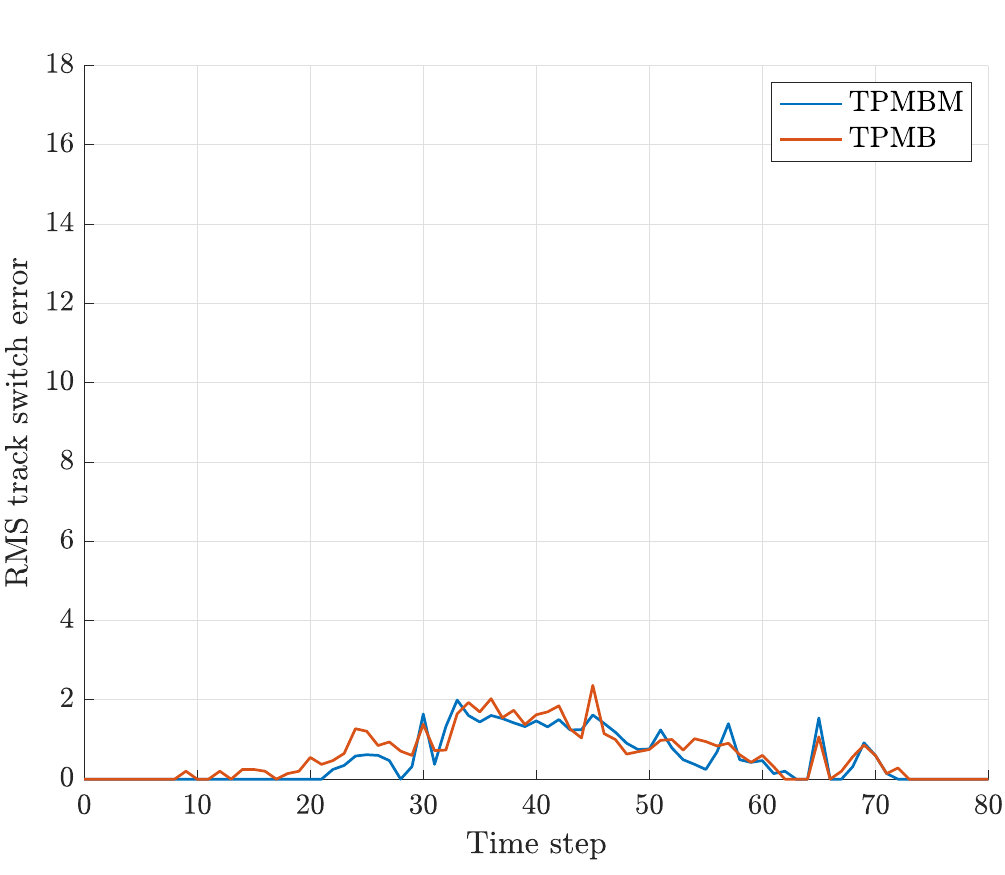}
    \caption{TGOSPA error and its decomposition.}
    \vspace{-4mm}
    \label{fig_tgospa}
\end{figure*}

\begin{figure*}[!t]
    \center
    \includegraphics[width=0.32\linewidth]{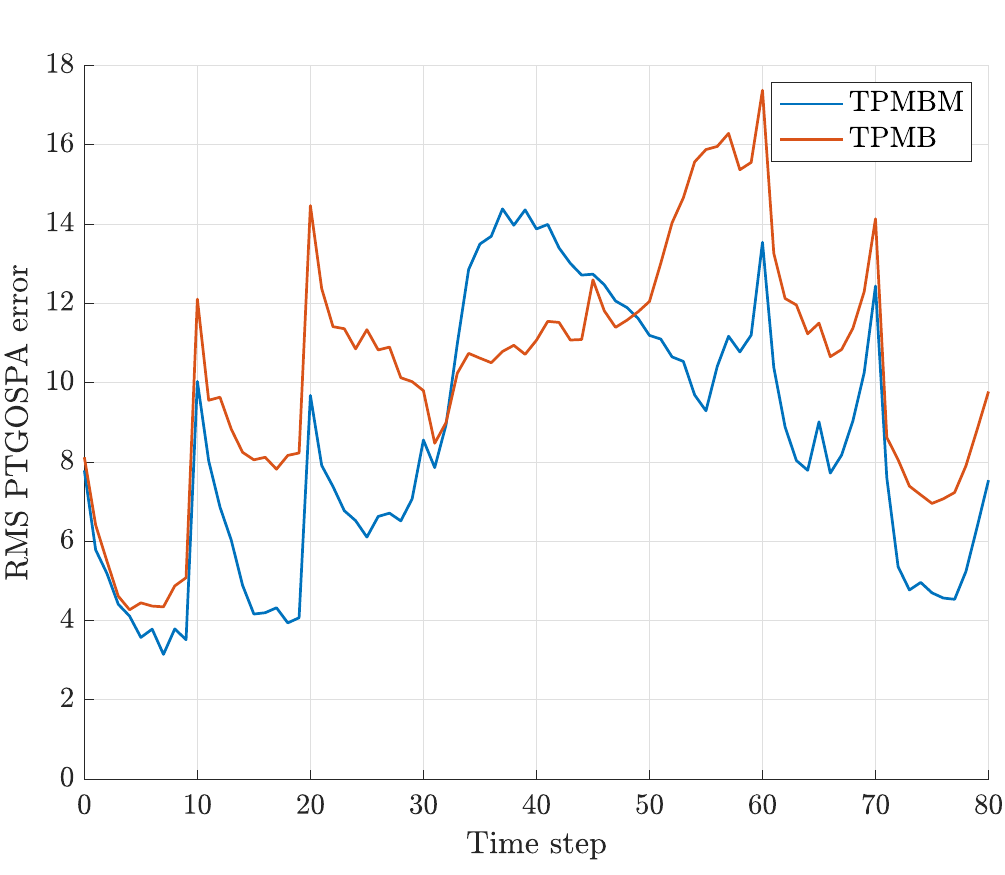}
    \includegraphics[width=0.32\linewidth]{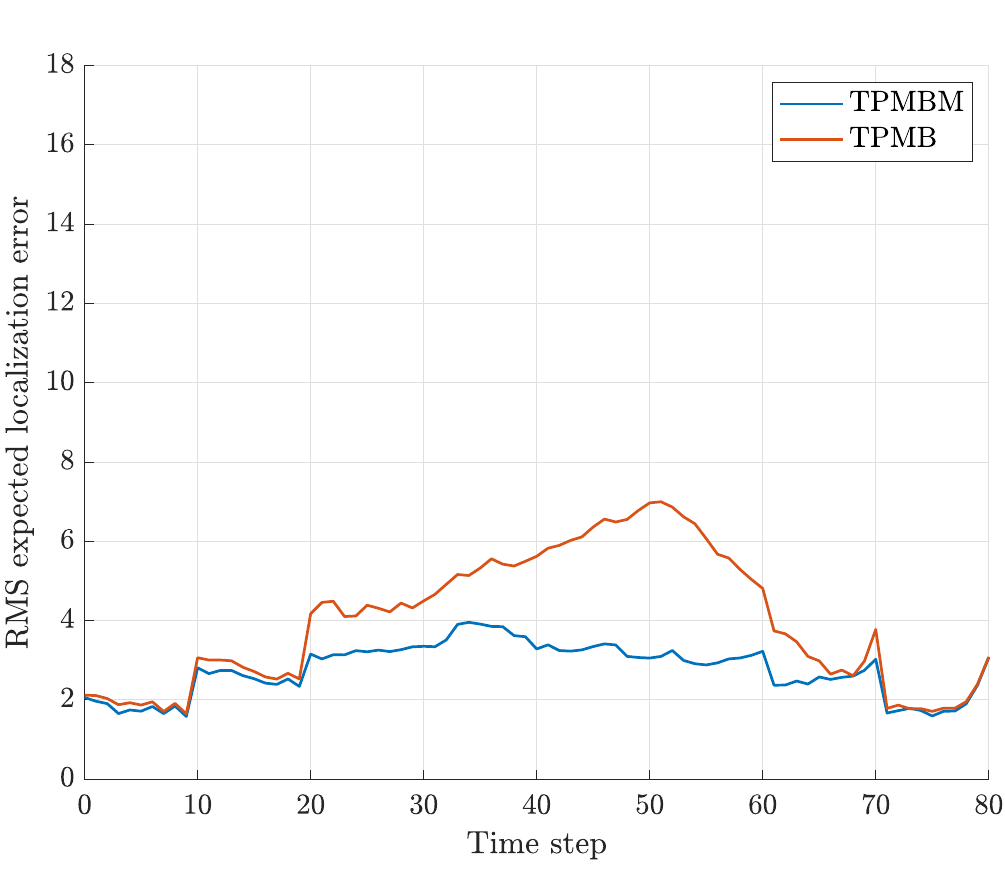}
    \includegraphics[width=0.32\linewidth]{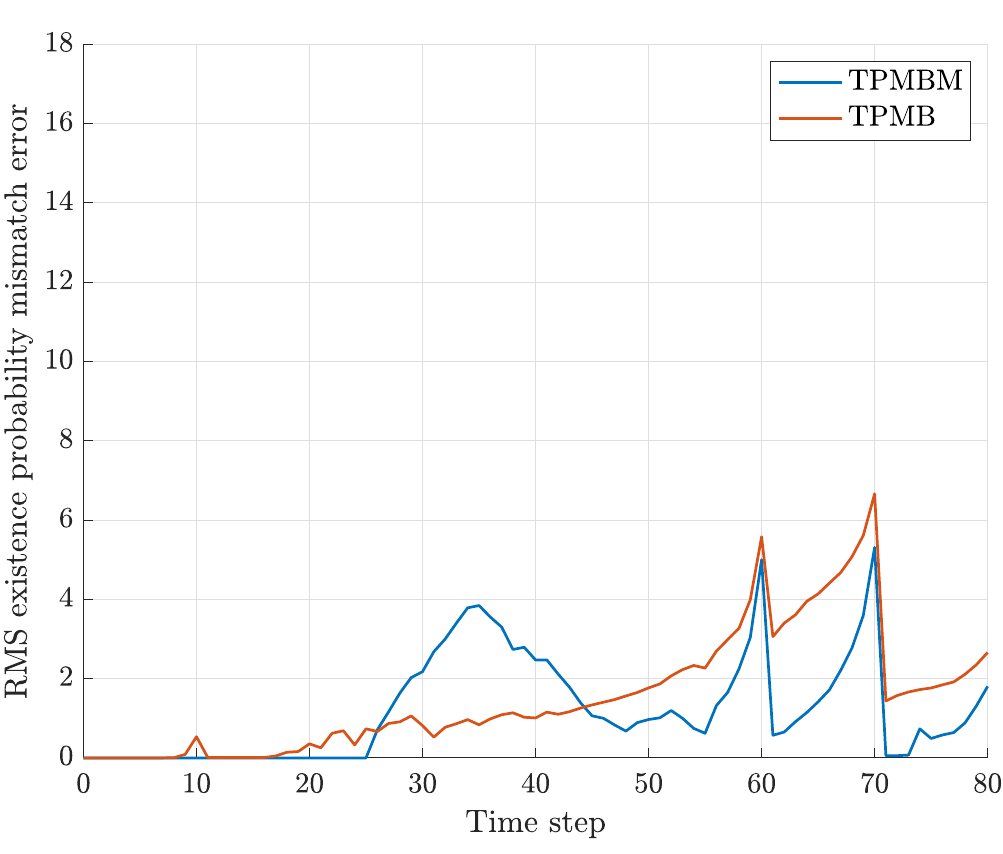}
    \includegraphics[width=0.32\linewidth]{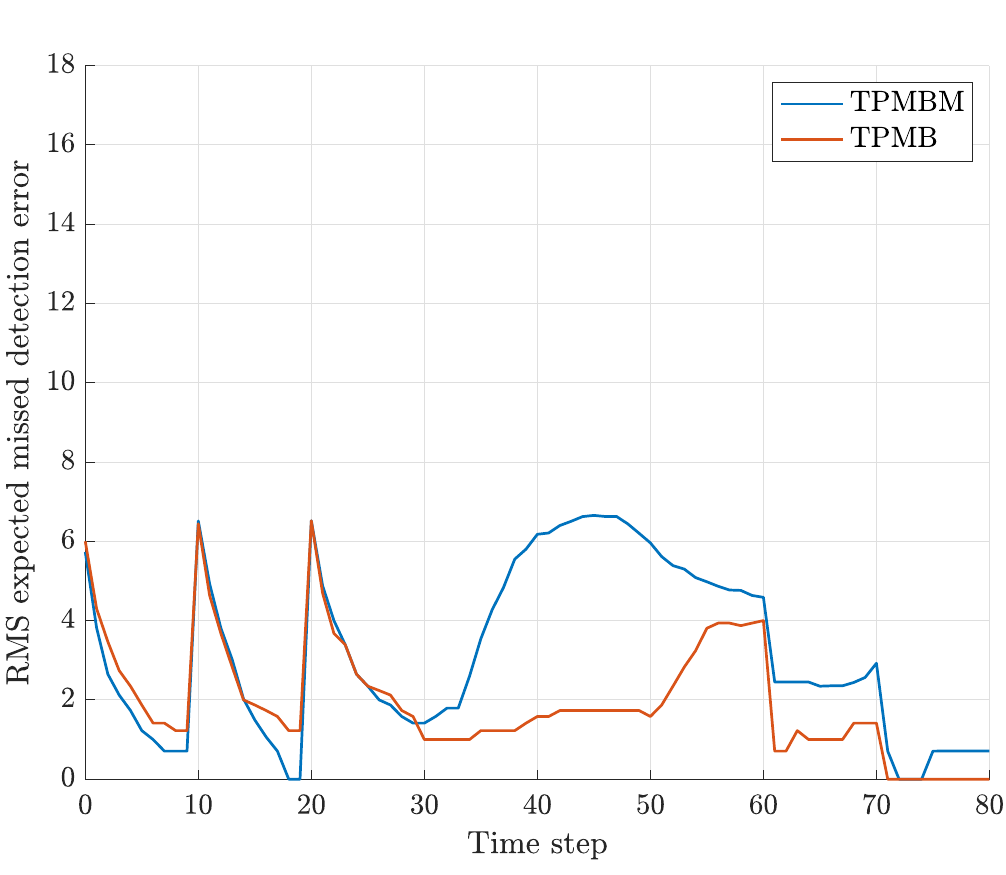}
    \includegraphics[width=0.32\linewidth]{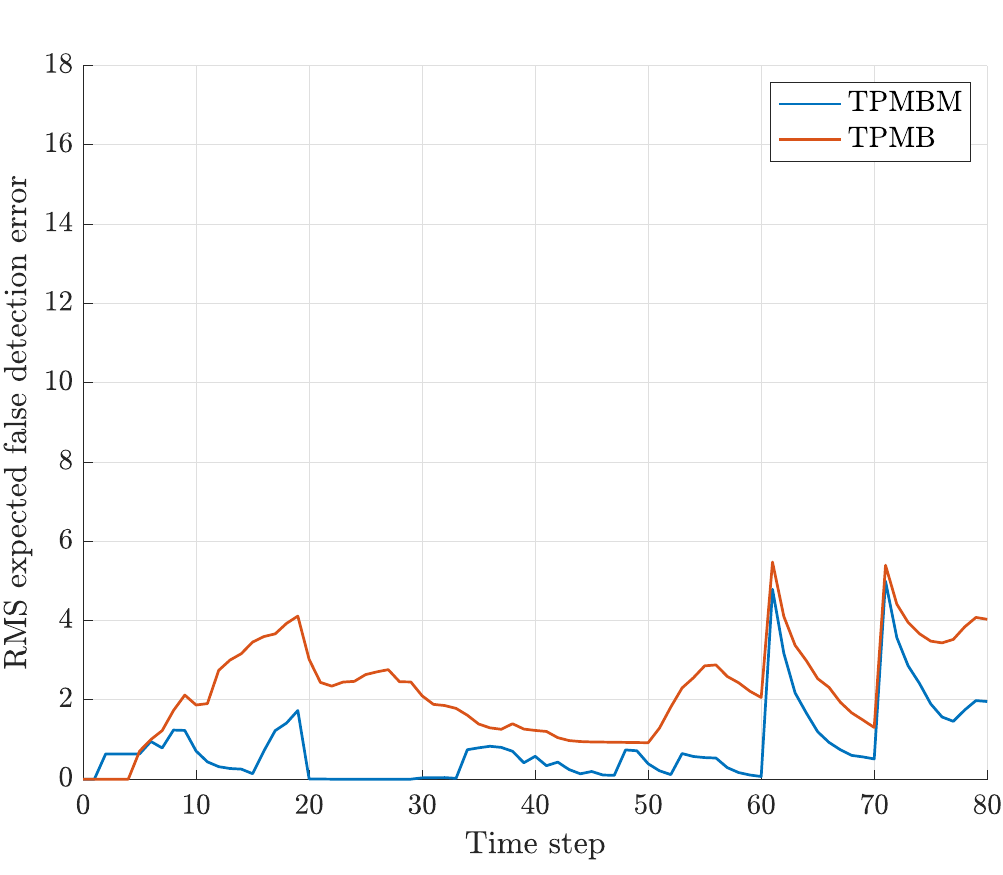}
    \includegraphics[width=0.32\linewidth]{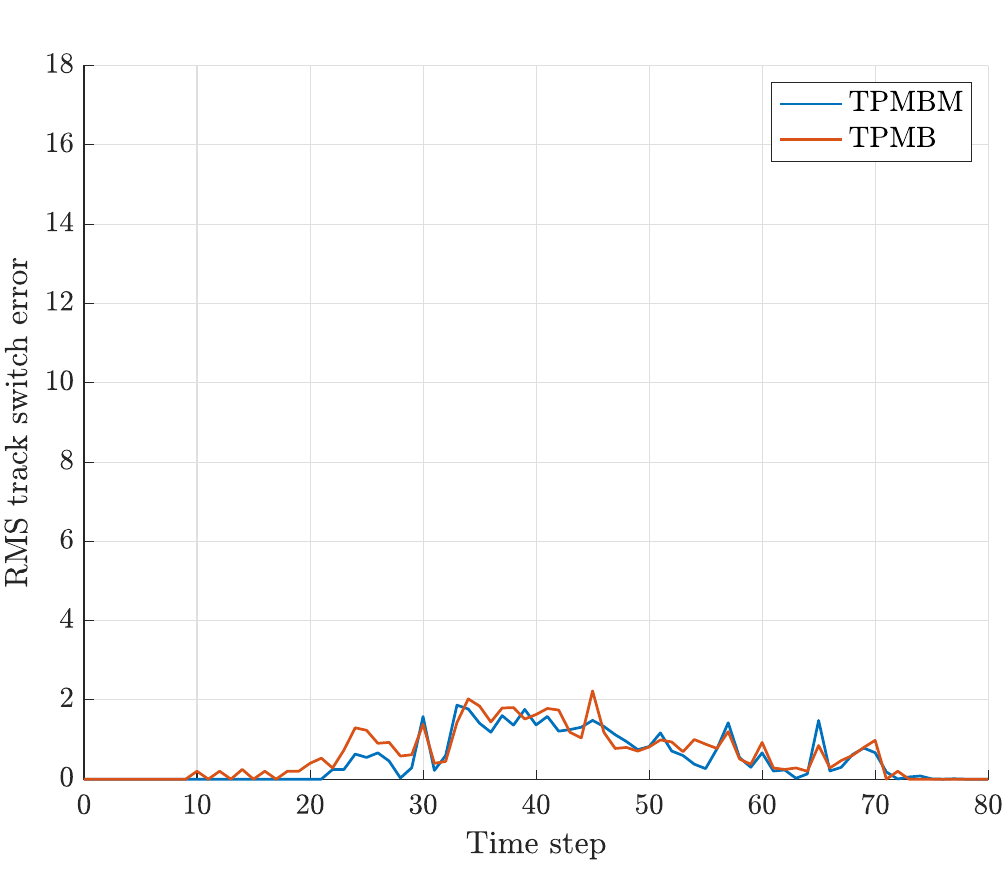}
    \caption{PTGOSPA error and its decomposition.}
    \label{fig_ptgospa}
    \vspace{-6mm}
\end{figure*}

We conduct 100 Monte Carlo simulations and compute average the root-mean-square (RMS) TGOSPA and PTGOSPA errors and their decompositions for each filter. The TGOSPA error and its decomposition over time are shown in Fig.~\ref{fig_tgospa}, whereas the PTGOSPA error and its decomposition over time are shown in Fig.~\ref{fig_ptgospa}. The results demonstrate that the trends of TGOSPA and PTGOSPA generally align, both increasing as objects come into close proximity, with notable changes in missed detection errors when objects are born and in false detections errors when objects become dead. 

The major difference lies in the missed detection error of the TGOSPA and PTGOSPA decompositions when objects become dead: which increases in TGOSPA but decreases in PTGOSPA and instead shows a rise in existence probability mismatch error. This suggests that, when objects disappear, the corresponding time steps do not coincide with the modes of their probability mass functions for being alive. However, under PTGOSPA, their marginal object states at corresponding time steps are still matched to the true objects, and the error is captured as a mismatch in existence probabilities.

\section{Conclusions}

In this paper, we have presented a metric for uncertainty-aware MOT performance evaluation, called PTGOSPA. It is a probabilistic generalization of the TGOSPA metric, extended to the space of sets of time sequences of Bernoulli densities through the use of the PGOSPA metric. We have also showed that PTGOSPA can be decomposed into five components: expected localization error and existence probability mismatch error for properly detected objects, expected missed and false detection error, and track switch error. 

Future work includes using PTGOSPA to quantify approximation errors in recursive MOT filtering based on sets of trajectories, analogous to the use of PGOSPA in \cite{xia2024probabilistic}, and extending PTGOSPA to incorporate time weighting schemes as in \cite{garcia2021time}.

\bibliographystyle{IEEEtran}
\bibliography{references.bib}

\begin{thebibliography}{10}
\providecommand{\url}[1]{#1}
\csname url@samestyle\endcsname
\providecommand{\newblock}{\relax}
\providecommand{\bibinfo}[2]{#2}
\providecommand{\BIBentrySTDinterwordspacing}{\spaceskip=0pt\relax}
\providecommand{\BIBentryALTinterwordstretchfactor}{4}
\providecommand{\BIBentryALTinterwordspacing}{\spaceskip=\fontdimen2\font plus
\BIBentryALTinterwordstretchfactor\fontdimen3\font minus \fontdimen4\font\relax}
\providecommand{\BIBforeignlanguage}[2]{{%
\expandafter\ifx\csname l@#1\endcsname\relax
\typeout{** WARNING: IEEEtran.bst: No hyphenation pattern has been}%
\typeout{** loaded for the language `#1'. Using the pattern for}%
\typeout{** the default language instead.}%
\else
\language=\csname l@#1\endcsname
\fi
#2}}
\providecommand{\BIBdecl}{\relax}
\BIBdecl

\bibitem{blackman1999design}
S.~S. Blackman and R.~Popoli, \emph{Design and analysis of modern tracking systems}.\hskip 1em plus 0.5em minus 0.4em\relax Artech House Publishers, 1999.

\bibitem{bernardin2008evaluating}
K.~Bernardin and R.~Stiefelhagen, ``Evaluating multiple object tracking performance: the clear {MOT} metrics,'' \emph{EURASIP Journal on Image and Video Processing}, vol. 2008, pp. 1--10, 2008.

\bibitem{luiten2021hota}
J.~Luiten, A.~Osep, P.~Dendorfer, P.~Torr, A.~Geiger, L.~Leal-Taix{\'e}, and B.~Leibe, ``{HOTA}: {A} higher order metric for evaluating multi-object tracking,'' \emph{International Journal of Computer Vision}, vol. 129, pp. 548--578, 2021.

\bibitem{ristic2011metric}
B.~Ristic, B.-N. Vo, D.~Clark, and B.-T. Vo, ``A metric for performance evaluation of multi-target tracking algorithms,'' \emph{IEEE Transactions on Signal Processing}, vol.~59, no.~7, pp. 3452--3457, 2011.

\bibitem{fridling1991performance}
B.~E. Fridling and O.~E. Drummond, ``Performance evaluation methods for multiple-target-tracking algorithms,'' in \emph{Signal and Data Processing of Small Targets 1991}, vol. 1481.\hskip 1em plus 0.5em minus 0.4em\relax SPIE, 1991, pp. 371--383.

\bibitem{drummond1992ambiguities}
O.~E. Drummond and B.~E. Fridling, ``Ambiguities in evaluating performance of multiple target tracking algorithms,'' in \emph{Signal and Data Processing of Small Targets 1992}, vol. 1698.\hskip 1em plus 0.5em minus 0.4em\relax SPIE, 1992, pp. 326--337.

\bibitem{schuhmacher2008new}
D.~Schuhmacher and A.~Xia, ``A new metric between distributions of point processes,'' \emph{Advances in Applied Probability}, vol.~40, no.~3, pp. 651--672, 2008.

\bibitem{schuhmacher2008consistent}
D.~Schuhmacher, B.-T. Vo, and B.-N. Vo, ``A consistent metric for performance evaluation of multi-object filters,'' \emph{IEEE Transactions on Signal Processing}, vol.~56, no.~8, pp. 3447--3457, 2008.

\bibitem{rahmathullah2017generalized}
A.~S. Rahmathullah, {\'A}.~F. Garc{\'i}a-Fern{\'a}ndez, and L.~Svensson, ``Generalized optimal sub-pattern assignment metric,'' in \emph{20th International Conference on Information Fusion (Fusion)}.\hskip 1em plus 0.5em minus 0.4em\relax IEEE, 2017, pp. 1--8.

\bibitem{garcia2019spooky}
{\'A}.~F. Garc{\'i}a-Fem{\'a}ndez and L.~Svensson, ``Spooky effect in optimal {OSPA} estimation and how {GOSPA} solves it,'' in \emph{22th International Conference on Information Fusion (FUSION)}.\hskip 1em plus 0.5em minus 0.4em\relax IEEE, 2019, pp. 1--8.

\bibitem{garcia2021analysis}
{\'A}.~F. Garc{\'i}a-Fern{\'a}ndez, M.~Hernandez, and S.~Maskell, ``An analysis on metric-driven multi-target sensor management: {GOSPA} versus {OSPA},'' in \emph{24th International Conference on Information Fusion (FUSION)}.\hskip 1em plus 0.5em minus 0.4em\relax IEEE, 2021, pp. 1--8.

\bibitem{beard2017ospa}
M.~Beard, B.~T. Vo, and B.-N. Vo, ``{OSPA}(2): {U}sing the {OSPA} metric to evaluate multi-target tracking performance,'' in \emph{2017 International Conference on Control, Automation and Information Sciences (ICCAIS)}.\hskip 1em plus 0.5em minus 0.4em\relax IEEE, 2017, pp. 86--91.

\bibitem{garcia2020metric}
{\'A}.~F. Garc{\'i}a-Fern{\'a}ndez, A.~S. Rahmathullah, and L.~Svensson, ``A metric on the space of finite sets of trajectories for evaluation of multi-target tracking algorithms,'' \emph{IEEE Transactions on Signal Processing}, vol.~68, pp. 3917--3928, 2020.

\bibitem{krejvci2024tgospa}
J.~Krej{\v{c}}{\'\i}, O.~Kost, O.~Straka, Y.~Xia, L.~Svensson, and {\'A}.~F. Garc{\'\i}a-Fern{\'a}ndez, ``{TGOSPA} metric parameters selection and evaluation for visual multi-object tracking,'' \emph{arXiv preprint arXiv:2412.08321}, 2024.

\bibitem{garcia2021time}
{\'A}.~F. Garc{\'i}a-Fern{\'a}ndez, A.~S. Rahmathullah, and L.~Svensson, ``A time-weighted metric for sets of trajectories to assess multi-object tracking algorithms,'' in \emph{24th International Conference on Information Fusion (FUSION)}.\hskip 1em plus 0.5em minus 0.4em\relax IEEE, 2021, pp. 1--8.

\bibitem{xia2024probabilistic}
Y.~Xia, {\'A}.~F. Garc{\'\i}a-Fern{\'a}ndez, J.~Karlsson, T.~Yuan, K.-C. Chang, and L.~Svensson, ``Probabilistic {GOSPA}: {A} metric for performance evaluation of multi-object filters with uncertainties,'' \emph{arXiv preprint arXiv:2412.11482}, 2024.

\bibitem{musicki2004joint}
D.~Musicki and R.~Evans, ``Joint integrated probabilistic data association: {JIPDA},'' \emph{IEEE Transactions on Aerospace and Electronic Systems}, vol.~40, no.~3, pp. 1093--1099, 2004.

\bibitem{reuter2014labeled}
S.~Reuter, B.-T. Vo, B.-N. Vo, and K.~Dietmayer, ``The labeled multi-{B}ernoulli filter,'' \emph{IEEE Transactions on Signal Processing}, vol.~62, no.~12, pp. 3246--3260, 2014.

\bibitem{garcia2019multiple}
{\'A}.~F. Garc{\'i}a-Fern{\'a}ndez, L.~Svensson, and M.~R. Morelande, ``Multiple target tracking based on sets of trajectories,'' \emph{IEEE Transactions on Aerospace and Electronic Systems}, vol.~56, no.~3, pp. 1685--1707, 2020.

\bibitem{granstrom2025poisson}
K.~Granstr{\"o}m, L.~Svensson, Y.~Xia, J.~Williams, and {\'A}.~F. Garc{\'\i}a-Fern{\'a}ndez, ``Poisson multi-{B}ernoulli mixtures for sets of trajectories,'' \emph{IEEE Transactions on Aerospace and Electronic Systems}, vol.~61, no.~2, pp. 5178--5194, 2025.

\bibitem{xia2019multi}
Y.~Xia, K.~Granstr{\"o}m, L.~Svensson, {\'A}.~F. Garc{\'i}a-Fern{\'a}ndez, and J.~L. Williams, ``Multi-scan implementation of the trajectory {P}oisson multi-{B}ernoulli mixture filter,'' \emph{Journal of Advances in Information Fusion}, vol.~14, no.~2, pp. 213--235, 2019.

\bibitem{garcia2020trajectory}
{\'A}.~F. Garc{\'i}a-Fern{\'a}ndez, L.~Svensson, J.~L. Williams, Y.~Xia, and K.~Granstr{\"o}m, ``Trajectory {P}oisson multi-{B}ernoulli filters,'' \emph{IEEE Transactions on Signal Processing}, vol.~68, pp. 4933--4945, 2020.

\bibitem{garcia2020trajectory2}
------, ``Trajectory multi-{B}ernoulli filters for multi-target tracking based on sets of trajectories,'' in \emph{23rd International Conference on Information Fusion (FUSION)}.\hskip 1em plus 0.5em minus 0.4em\relax IEEE, 2020, pp. 1--8.

\bibitem{mahler2007statistical}
R.~P. Mahler, \emph{{Statistical Multisource-Multitarget Information Fusion}}.\hskip 1em plus 0.5em minus 0.4em\relax Artech House, 2007.

\bibitem{crouse2016implementing}
D.~F. Crouse, ``On implementing 2{D} rectangular assignment algorithms,'' \emph{IEEE Transactions on Aerospace and Electronic Systems}, vol.~52, no.~4, pp. 1679--1696, 2016.

\bibitem{xia2024markov}
Y.~Xia, {\'A}.~F. Garc{\'\i}a-Fern{\'a}ndez, and L.~Svensson, ``Markov chain {Monte Carlo} multiscan data association for sets of trajectories,'' \emph{IEEE Transactions on Aerospace and Electronic Systems}, vol.~60, no.~6, pp. 7804--7819, 2024.

\end{thebibliography}

\end{document}